\documentclass{article}

\pdfoutput=1

\usepackage{hyperref, graphicx, amsmath, amssymb, authblk}
\usepackage[margin=1in]{geometry}
\usepackage[numbers]{natbib}
\title{Logistic push: a regression framework for partial AUC optimization}

\begin{document}

\author[1]{Travis A.\ Gerke\thanks{travis.gerke@moffitt.org}}
\author[2,3]{Svitlana Tyekucheva}
\author[4]{Lorelei A.~Mucci}
\author[2,3]{Giovanni Parmigiani}

\affil[1]{{\footnotesize Department of Cancer Epidemiology, Moffitt Cancer Center, Tampa, FL, USA}}
\affil[2]{{\footnotesize Department of Biostatistics and Computational Biology, Dana-Farber Cancer Institute, Boston, MA, USA}}
\affil[3]{{\footnotesize Department of Biostatistics, Harvard School of Public Health, Boston, MA, USA}}
\affil[4]{{\footnotesize Department of Epidemiology, Harvard School of Public Health, Boston, MA, USA}}

\maketitle

\begin{abstract}
The area under the receiver operating characteristic curve (AUC) is often used to evaluate the performance of clinical prediction models. Recently, a more refined strategy has been proposed to examine a partial area under the curve (pAUC), which can account for differing costs associated with false negative versus false positive results. Such consideration can substantially increase the clinical utility of prediction models depending on the clinical question. Properties of the pAUC estimator create significant challenges for pAUC-optimal marker selection and model building. As such, current approaches towards these aims can be complex and computationally intensive. We present a simpler method based on weighted logistic regressions. We refer to our strategy as logistic push, due to shared heuristics with the ranking algorithm P-norm push. Logistic push is particularly useful in the high-dimensional setting, where fast and broadly available algorithms for fitting penalized regressions can be used for both marker selection and model fitting.
\end{abstract}

\section{Introduction}
Considerable effort in clinical research is devoted to the development of accurate tools that can guide clinical decision-making by predicting binary outcomes. Successful clinical or biological markers for prediction, proposed as singletons or in a multivariable signature, have the potential to improve screening practices, reveal likely disease prognoses, or identify those patients who would benefit from treatment. Commonly, the classification performance of a marker-based score is assessed by testing whether the area under an ROC curve (AUC) is improved when the score replaces or is added to an existing model\cite{hanley1983}. The AUC aggregates information regarding the sensitivity and specificity of a continuous classifier over all possible cutpoints, and can be interpreted as the probability that a randomly selected diseased patient will have a higher predicted risk than that of a randomly selected non-diseased patient. In this sense, the AUC provides a simple summary measure of overall performance.

\begin{figure}[!htb]
\centering
\includegraphics[width=\linewidth, trim=0cm 0cm 1cm 0cm]{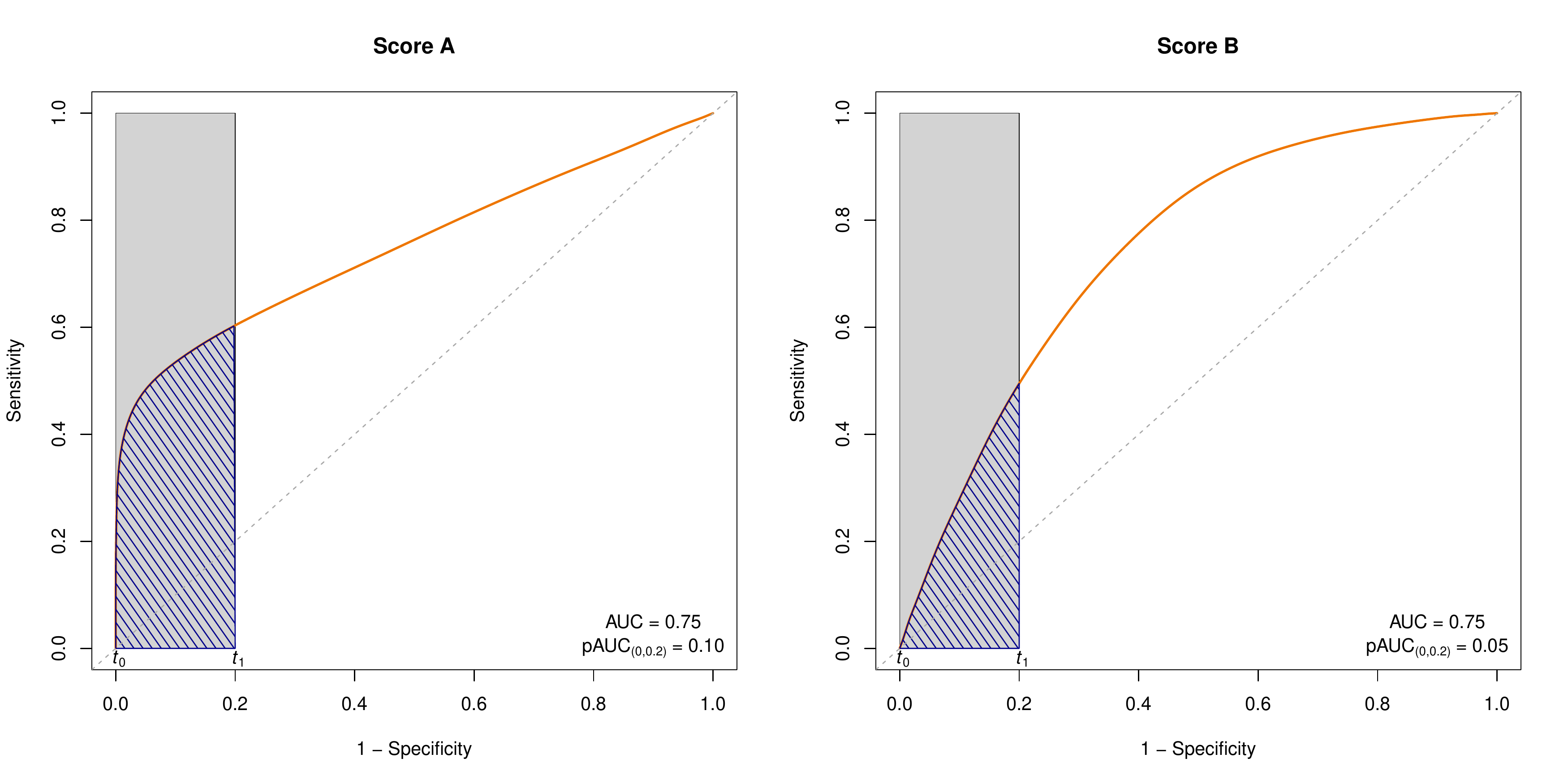}
\caption{\small ROC curves for two simulated continuous classifiers, Score A and Score B. Each has the same AUC (0.75), however, if the clinical cost of false positives is high, Score A is to be preferred due to its superior pAUC.}
\label{paucEx}
\end{figure}

Depending on the clinical question, the costs associated with a false negative and false positive result may not be the same. Such costs depend on a balance of the clinical consequences of each type of result.  As such, a more refined strategy has gained popularity in recent years which, instead of calculating the area under the full ROC curve, computes the area under a clinically relevant subset of the curve, yielding the partial AUC (pAUC) statistic\cite{mcclish1989, thompson1989}. For example, certain cancer screening applications demand high specificity (e.g.~greater than 80\%) to prevent unnecessary biopsies\cite{baker2001}, rendering an analysis of the full AUC inappropriate. The need for high specificity is particularly pronounced in studies of rare diseases, where tests of inadequate specificity can lead to dramatic numbers of false positive results\cite{wang2011}. In these cases, an analysis of pAUC over a specificity range such as (0.80, 1.00) should be undertaken (Figure~\ref{paucEx}). Analogous reasoning can be used to investigate markers that discriminate well over high sensitivity regions\cite{jiang1996}.

Many statistical methods now permit analysis and comparison of pAUC for a selected marker set\cite{cai2008, dodd2003, ma2013, walter2005}. A more challenging task, however, concerns marker discovery and combination by way of pAUC optimization. This important extension has been discussed in the context of microarray experiments\cite{pepe2003, komori2010, schmid2012, wang2011}, but efforts have been hindered by the complex mathematical nature of the pAUC estimator\cite{hsu2013}, which is often approximated to arrive at a differentiable form for usability\cite{wang2011, ma2005, schmid2012}. The resulting algorithms are computationally burdensome, and may only produce a local solution, potentially failing to detect clinically important biomarkers\cite{hsu2013}.

In parallel, penalized regression techniques, such as the lasso\cite{tibshirani1994} and elastic net\cite{zou2005}, have enjoyed great success with respect to high-dimensional predictor selection with accompanying model development. Computationally efficient algorithms\cite{friedman2010,efron2004} have made it possible to fit models from high-dimensional data in seconds, leading to vast popularity and widespread availability. Although the objective function to be maximized under a penalized logistic regression is a statistical likelihood, not the AUC, the resulting models can often discriminate binary outcomes with appreciable accuracy.

Given the speed and ease of use of the penalized regression approach, an alternative method for finding pAUC-optimal marker combinations that leverages the regression algorithms may provide a valuable resource. Connections between full AUC and logistic regression have been discussed previously\cite{pepe2005, pepe2006}, however, implications for high-dimensional settings and pAUC were not investigated. In this paper, we establish a relationship between pAUC and logistic regression that gives rise to a computationally efficient method for pAUC-based marker selection and model fitting. 

\section{Partial area under the curve}
Suppose we have $i=1,\ldots,n$ sample vectors of the form $[\begin{matrix}y_i&x_{i1}&\cdots&x_{ip}\end{matrix}]$, where $y_i\in\{-1,\,1\}$ is a binary outcome variable and $x_{i_1},\ldots,x_{ip}\in\mathbb{R}$ are marker variables. Let subscripts $j=1,\ldots,J$ denote subjects from the diseased group and $k=1,\ldots,K$ be those from the non-diseased group, with ${J+K=n}$. Generally, we wish to derive a continuous classifier $f:\mathbb{R}^p\to\mathbb{R}$ which is accurate in the sense that ${f(x_j)>f(x_k)}$ for a clinically important set of $j\times k$ pairings.

For a given cutpoint $c$, test sensitivity is defined as the probability of a true positive classification, $P[f(X_j)>c]$. Similarly, the quantity ($1-\text{specificity}$), commonly called the false positive rate (FPR), is the probability of a false positive classification, $P[f(X_k)>c]$. The receiver operating characteristic (ROC) curve plots sensitivity versus ($1-\text{specificity}$) for all possible cutpoints. The AUC is the area under the ROC curve, and has the useful probabilistic interpretation of $P[f(X_j)>f(X_k)]$.

The pAUC over FPR interval $(t_0,\,t_1)$ is defined as 
\begin{align}\label{paucDef}
\text{pAUC}(t_0,\,t_1)=\int_{t_0}^{t_1}\text{ROC}(t)\,dt.
\end{align}
Thus, pAUC provides a measure of classifier sensitivity over a predetermined range of clinically acceptable false positive rates. Note that, for the case $(t_0,\, t_1)=(0,\,1)$, pAUC is equivalent to the full AUC. For ease of exposition, and because interest typically lies in the left-most portion of the ROC curve, we restrict our attention to FPR ranges of the form $(0,\,t)$ with corresponding partial area denoted $\text{pAUC}_t$. 

Dodd and Pepe\cite{dodd2003} provide a simple non-parametric estimator of pAUC$_t$ as
\begin{align}\label{paucEst}
\widehat{\text{pAUC}_t}=\frac{1}{JK}\sum_{j=1}^{J}\sum_{k=1}^{K}\mathbf{1}[f(x_j)>f(x_k),\,f(x_k)>q^-_{(1-t)}],
\end{align}
where $q^-_{(1-t)}$ represents the $(1-t)$ quantile of classifier scores in the non-diseased population. Largely owing to the non-differentiability of the indicator function, maximization of this objective with respect to $f$ is difficult, particularly when many markers comprise the risk score or when variable selection must be performed\cite{pepe2006}. Some existing methods approximate the indicator using smooth functions, after which a boosting or wrapper-style algorithm is implemented\cite{komori2010,schmid2012,wang2011}. Other approaches implement a support vector method to minimize a convex upper bound on the pAUC loss function\cite{narasimhan2013,narasimhan2014}.

When the left-most region of the ROC curve is of interest, the problem of pAUC maximization can be recast as a problem of ranking at the top of the list. To see this, observe that maximization of the estimator in (\ref{paucEst}) is equivalent to minimization with respect to $f$ of 
\begin{align}\label{cost1}
L_{(\text{0-1})}=\sum_{k:\,f(x_k)>q^-_{(1-t)}}\left(\sum_{j=1}^{J}\mathbf{1}[f(x_j)<f(x_k)]\right).
\end{align}
This formulation reveals that non-diseased subjects with low predicted risk scores (i.e.~${f(x_k)<q^-_{(1-t)}}$) do not contribute to the quantification of error, while each non-diseased subject with a high risk score contributes to the outer sum a quantity equal to the number of diseased subjects with a lower score. Hence, increases in pAUC$_t$ are attained by ``pushing'' high-scoring non-diseased subjects down on the list of ranked scores.

Rudin\cite{rudin2009} proposed P-norm push as a general method to minimize a closely related class of functions, 
\begin{align}\label{costp}
L_p=\sum_{k=1}^K\left(\sum_{j=1}^{J}\mathbf{1}[f(x_j)<f(x_k)]\right)^p.
\end{align}
Here, $p$ determines the price for high-scoring non-diseased subjects, with larger $p$ corresponding to greater penalties for those appearing at the top of the ranked list. Minimization of a convex upper bound for (\ref{costp}) proceeds by way of a boosting algorithm. P-norm push is not explicitly designed to optimize pAUC$_t$, although as $p\to\infty$, the so-called infinite push\cite{agarwal2011} can be seen as a maximizer of pAUC$_{1/K}$\cite{narasimhan2013}.

\section{The logistic push method}
A standard strategy for evaluating associations between a set of markers and a binary outcome is to fit a logistic regression. The objective to be maximized is a logistic likelihood, and $f$ is estimated through minimization of
\begin{align}\label{costL}
L_{\mathcal L}=\sum_{j=1}^J\log\left\{1+\exp\left[-f\left(x_j\right)\right]\right\}+\sum_{k=1}^K\log\left\{1+\exp\left[f\left(x_k\right)\right]\right\}.
\end{align}
It is straightforward to show that $L_{\mathcal L}/\log(2)$ forms an upper bound for 
\begin{align}\label{costL01}
L_{\mathcal L\text{(0-1)}}=\sum_{j=1}^J\mathbf{1}[f(x_j)<0]+\sum_{k=1}^K\mathbf{1}[f(x_k)>0],
\end{align}
implying that logistic regression minimization of (\ref{costL}) can lead to reductions in (\ref{costL01}). 

Recall the heuristic link between P-norm push and the quantity in (\ref{cost1}) which tells us that, in order to maximize pAUC$_t$, we wish to push non-diseased subjects down from the top of the ranked score list. To achieve this goal, we can modify (\ref{costL01}) as
\begin{align}\label{costL01mod}
L_{\mathcal L\text{(0-1)}}^*=\sum_{j=1}^J\mathbf{1}[f(x_j)<q^-_{(1-t)}]+\sum_{k=1}^Kw\mathbf{1}[f(x_k)>q^-_{(1-t)}],
\end{align}
where $w\ge1$ is a weight that determines the magnitude of penalty assigned to high-ranking non-diseased predictions. For example, the use of a large $w$ will impose a strict penalty for non-diseased sample predictions that fall above the population quantile $\smash{q^-_{(1-t)}}$, and the resulting estimate of $f$ will aim to produce few such occurrences. Relating $L_{\mathcal L\text{(0-1)}}^*$ back to (\ref{costL}), we now might wish to minimize
\begin{align}\label{costLmod}
L_{\mathcal L}^*=\sum_{j=1}^J\log\left\{1+\exp\left[-\left(f\left(x_j\right)-q^-_{(1-t)}\right)\right]\right\}+\sum_{k=1}^Kw\log\left\{1+\exp\left[f\left(x_k\right)-q^-_{(1-t)}\right]\right\}.
\end{align}
Of course, because pAUC$_t$ and its associated objective in (\ref{cost1}) are rank-based measures, the offset imposed by $\smash{q^-_{(1-t)}}$ has no effect. Hence, we treat the term as a nuisance parameter, and conveniently avoid the problem of $q^-_{(1-t)}$ estimation which is a necessary step in related methods\cite{pepe2000}. The resulting function to be minimized through the logistic push method is
\begin{align}\label{costLpush}
L_{\text{lp}}=\sum_{j=1}^J\log\left\{1+\exp\left[-f\left(x_j\right)\right]\right\}+\sum_{k=1}^K\hat w\log\left\{1+\exp\left[f\left(x_k\right)\right]\right\},
\end{align}
which simply gives rise to a weighted logistic regression. The weight $\hat w$ given to non-diseased subjects is considered a tuning parameter, and can be chosen through cross-validation with $\widehat{\text{pAUC}}_t$ as the objective. 

When a high-dimensional candidate set of markers is available, a further term can be added to (\ref{costLpush}) to implement a penalized logistic regression\cite{hastie2009}. For example, common choices include $\lambda_1\lVert\boldsymbol\beta\rVert_1$ (lasso), $\lambda_2\lVert\boldsymbol\beta\rVert_2$ (ridge), or $\lambda_1\lVert\boldsymbol\beta\rVert_1 + \lambda_2\lVert\boldsymbol\beta\rVert_2$ (elastic net). Appropriate choice of the penalty term will depend on the clinical application and the desired level of model sparsity. For simplicity in what follows, we constrain our attention to the logistic lasso, and consider linear predictor specifications of the form $f(\mathbf x)=\beta_0+\boldsymbol\beta^T\mathbf x$, where $\boldsymbol\beta$ and $\mathbf x$ are $(p\times 1)$ vectors.

\section{Performance of logistic push}
Using simulations we compared logistic push to the standard logistic lasso and a pAUC boosting technique, pAUC-GBS\cite{schmid2012}, which is representative of similar methods in that sigmoid smoothers are used to approximate indicator functions. Two performance measures are of interest: (i) whether logistic push is able to select markers that increase pAUC$_t$ and (ii) whether logistic push combines selected markers in a way that pAUC$_t$ is high.

The same simulated data used by Schmid et al.\cite{schmid2012} to show the efficacy of pAUC-GBS was analyzed in our comparative study. Each of 100 independent data sets consisted of 50 diseased and 50 non-diseased subjects with 506 candidate variables denoted $X_1,\ldots,X_{506}$. Markers $X_1$, $X_2$, and $X_3$ were generated under the distribution shown by Score A in Figure~\ref{paucEx}. Similarly, markers $X_4$, $X_5$, and $X_6$ were generated as Score B. The 500 markers $X_7,\ldots.X_{506}$ were drawn as non-informative noise. Thus, a strong pAUC$_t$ strategy for small $t$ will only select markers $X_1$, $X_2$, and $X_3$ and combine them in a way that high pAUC$_t$ is realized. 

We assessed performance for classifier specificities of at least 80\%, corresponding to pAUC$_{0.20}$. Stratified 5-fold cross-validation (CV) was used to select the weights $\hat w$ for logistic push, with nested stratified 5-fold CV used to select the lasso penalty $\lambda$. Estimated $\widehat{\text{pAUC}}_{0.20}$ was used as the objective for both CV procedures. The settings used for pAUC-GBS were the same as those described in the appendix of Schmid et al.\cite{schmid2012} To fit the standard lasso regressions, default settings in the R package \verb|glmnet| were used under 5-fold CV for $\lambda$ selection with logistic deviance as the objective. A linear predictor was specified for all three approaches. For each of the 100 iterations, we simulated an i.i.d.~test data set to evaluate external pAUC$_t$ performance. 

\begin{figure}[!htb]
\centering
\includegraphics[width=.9\linewidth]{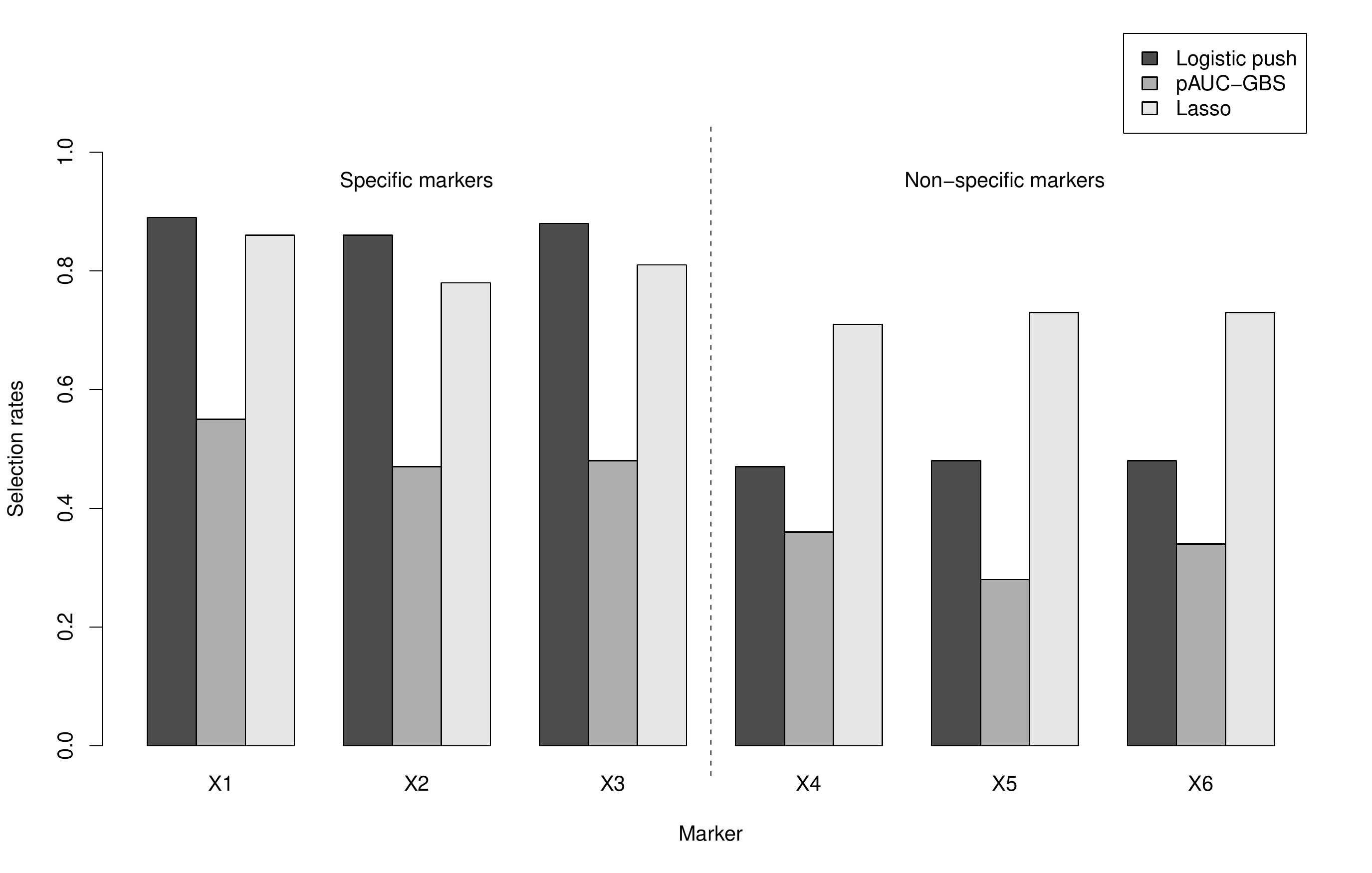}
\caption{\small Selection rates using logistic push, pAUC-GBS, and the lasso for the 6 informative markers in the simulation study.}
\label{selection}
\end{figure}

Selection performance for the 6 informative markers is summarized in Figure~\ref{selection}. Logistic push outperforms both pAUC-GBS and the lasso in preferentially selecting the desired specific markers. Selection rates for the 500 noise variables were acceptably low with logistic push. Any given noise variable was selected an average of 0.93 times over of the 100 iterations, compared to 0.05 and 2.50 by the respective pAUC-GBS and lasso methods. 

Test data performance measured by pAUC$_{0.20}$ for models trained using logistic push was superior to the performance of both pAUC-GBS and the lasso. The average external estimate of pAUC$_{0.20}$ with logistic push was 0.124 ($\text{median}=0.124$; $\text{IQR}=[0.114,\,0.135]$), compared to 0.117 ($\text{median}=0.117$; $\text{IQR}=[0.108,\,0.128]$) for pAUC-GBS and 0.118 ($\text{median}=0.124$; $\text{IQR}=[0.105,\,0.132]$) for the lasso.

\section{Discussion}
In this paper, we have provided a justification for using a simple weighted logistic regression approach for building models that optimize pAUC, and have shown that this method is able to outperform the pAUC-GBS boosting technique which is designed to maximize pAUC. This result was somewhat surprising, given that pAUC-GBS is constructed to optimize pAUC$_t$ directly, while logistic push only seeks to minimize an upper bound on the associated loss function. We expect that pAUC-GBS could benefit from CV tuning for the $\sigma$ parameter in the sigmoid approximation to the indicator; however, such an approach would prove computationally infeasible once the number of candidate variables becomes too large. By contrast, logistic push utilizes lasso algorithms which enjoy much lower computational complexity for large $p$ problems compared to boosting algorithms. 

The weight $\hat w$ assigned to the non-diseased subjects in logistic push is a tuning parameter that we selected through cross-validation. The average weight selected in our simulation study was 8.88 ($\text{median}=9.50$; $\text{IQR}=[3.75,\,13.5]$). A more deterministic method for selecting this parameter could lead to gains in accuracy and efficiency, and future work will investigate this possibility. Interestingly, in the limit as $\hat w\to\infty$, we arrive at infinitely imbalanced logistic regression, for which resulting parameter estimates have been shown to have useful properties\cite{owen2007}. In particular, although $\beta_0\to-\infty$, probability ratios converge to the nontrivial quantity
\begin{align}\label{owen}
\frac{P(Y=1\mid X=x_1)}{P(Y=1\mid X=x_0)}\to\exp[f(x_1)-f(x_0)].
\end{align}
Functions of the form $g[f(x_j)-f(x_k)]$ are useful in the context of AUC ranking metrics, and were used by Rudin\cite{rudin2009}  to show upper bounds for ranking at the top of the list. 

If very few candidate variables are of interest, algorithms based on direct maximization of pAUC by way of grid search may exhibit superior accuracy\cite{pepe2006}. Of course, computational complexity of a grid search quickly becomes too great, and logistic push is intended to fill this gap. As such, the method is well-suited for marker discovery, though results from our simulation study indicate that the models produced by logistic push may reveal external pAUC performance that is comparable to methods which maximize pAUC directly. 

Many current efforts in biomarker research concern the search for prognostic indicators that exhibit either high specificity or high sensitivity. As an example of the former, contemporary clinical markers of aggressive prostate cancer, such as Gleason score, have high sensitivity but lack specificity\cite{andren2006}. Because most prostate cancer diagnoses are non-aggressive, the resulting positive predictive values of tests available to clinicians to guide treatment decisions are low, and the corresponding overtreatment rates are high\cite{loeb2014}. Hence, the availability of an efficient algorithm such as logistic push that can seek complementary markers of high specificity has the potential to substantially reduce such overtreatment and improve clinical decision-making.

\clearpage 

\section*{References}

\bibliography{pauc}

\begin{thebibliography}{30}
\providecommand{\natexlab}[1]{#1}
\providecommand{\url}[1]{\texttt{#1}}
\expandafter\ifx\csname urlstyle\endcsname\relax
  \providecommand{\doi}[1]{doi: #1}\else
  \providecommand{\doi}{doi: \begingroup \urlstyle{rm}\Url}\fi

\bibitem[Agarwal(2011)]{agarwal2011}
S.~Agarwal.
\newblock {T}he {I}nfinite {P}ush: a new support vector ranking algorithm that
  directly optimizes accuracy at the absolute top of the list.
\newblock In \emph{Proceedings of the SIAM International Conference on Data
  Mining}, pages 839--850, 2011.

\bibitem[Andren et~al.(2006)Andren, Fall, Franzen, Andersson, Johansson, and
  Rubin]{andren2006}
O.~Andren, K.~Fall, L.~Franzen, S.~O. Andersson, J.~E. Johansson, and M.~A.
  Rubin.
\newblock {{H}ow well does the {G}leason score predict prostate cancer death?
  {A} 20-year followup of a population based cohort in {S}weden}.
\newblock \emph{J. Urol.}, 175\penalty0 (4):\penalty0 1337--1340, Apr 2006.

\bibitem[Baker and Pinsky(2001)]{baker2001}
S.~G. Baker and P.~F. Pinsky.
\newblock {{A} proposed design and analysis for comparing digital and analog
  mammography: special receiver operating characteristic methods for cancer
  screening}.
\newblock \emph{J Am Statist Assoc}, 96\penalty0 (454):\penalty0 421--428, Jun
  2001.

\bibitem[Cai and Dodd(2008)]{cai2008}
T.~Cai and L.~E. Dodd.
\newblock {{R}egression analysis for the partial area under the {ROC} curve}.
\newblock \emph{Stat Sinica}, 18:\penalty0 817--836, 2008.

\bibitem[Dodd and Pepe(2003)]{dodd2003}
L.~E. Dodd and M.~S. Pepe.
\newblock {{P}artial {A}{U}{C} estimation and regression}.
\newblock \emph{Biometrics}, 59\penalty0 (3):\penalty0 614--623, Sep 2003.

\bibitem[Efron et~al.(2004)Efron, Hastie, Johnstone, and Tibshirani]{efron2004}
B.~Efron, T.~Hastie, I.~Johnstone, and R.~Tibshirani.
\newblock {{L}east angle regression}.
\newblock \emph{Ann Stat}, 32:\penalty0 407--499, 2004.

\bibitem[Friedman et~al.(2010)Friedman, Hastie, and Tibshirani]{friedman2010}
J.~Friedman, T.~Hastie, and R.~Tibshirani.
\newblock Regularization paths for generalized linear models via coordinate
  descent.
\newblock \emph{J Stat Softw}, 33\penalty0 (1):\penalty0 1--22, 2010.

\bibitem[Hanley and McNeil(1983)]{hanley1983}
J.~A. Hanley and B.~J. McNeil.
\newblock {{A} method of comparing the areas under receiver operating
  characteristic curves derived from the same cases}.
\newblock \emph{Radiology}, 148\penalty0 (3):\penalty0 839--843, Sep 1983.

\bibitem[Hastie et~al.(2009)Hastie, Tibshirani, and Friedman]{hastie2009}
T.~Hastie, R.~Tibshirani, and J.~Friedman.
\newblock \emph{The elements of statistical learning: Data mining, inference,
  and prediction}.
\newblock Springer, second edition, 2009.

\bibitem[Hsu and Hsueh(2013)]{hsu2013}
M.~Hsu and H.~Hsueh.
\newblock {{T}he linear combinations of biomarkers which maximize the partial
  area under the {ROC} curves}.
\newblock \emph{Comput Stat}, 28\penalty0 (2):\penalty0 647--666, Apr 2013.

\bibitem[Jiang et~al.(1996)Jiang, Metz, and Nishikawa]{jiang1996}
Y.~Jiang, C.~E. Metz, and R.~M. Nishikawa.
\newblock {{A} receiver operating characteristic partial area index for highly
  sensitive diagnostic tests}.
\newblock \emph{Radiology}, 201\penalty0 (3):\penalty0 745--750, Dec 1996.

\bibitem[Komori and Eguchi(2010)]{komori2010}
O.~Komori and S.~Eguchi.
\newblock {{A} boosting method for maximizing the partial area under the
  {R}{O}{C} curve}.
\newblock \emph{BMC Bioinformatics}, 11:\penalty0 314, 2010.

\bibitem[Loeb et~al.(2014)Loeb, Bjurlin, Nicholson, Tammela, Penson, Carter,
  Carroll, and Etzioni]{loeb2014}
S.~Loeb, M.A. Bjurlin, J.~Nicholson, T.L. Tammela, D.F. Penson, H.B. Carter,
  P.~Carroll, and R.~Etzioni.
\newblock {Overdiagnosis and overtreatment of prostate cancer}.
\newblock \emph{Eur. Urol.}, epub ahead of print, Jan 2014.

\bibitem[Ma et~al.(2013)Ma, Bandos, Rockette, and Gur]{ma2013}
H.~Ma, A.~I. Bandos, H.~E. Rockette, and D.~Gur.
\newblock {{O}n use of partial area under the {R}{O}{C} curve for evaluation of
  diagnostic performance}.
\newblock \emph{Stat Med}, 32\penalty0 (20):\penalty0 3449--3458, Sep 2013.

\bibitem[Ma and Huang(2005)]{ma2005}
S.~Ma and J.~Huang.
\newblock {{R}egularized {R}{O}{C} method for disease classification and
  biomarker selection with microarray data}.
\newblock \emph{Bioinformatics}, 21\penalty0 (24):\penalty0 4356--4362, Dec
  2005.

\bibitem[McClish(1989)]{mcclish1989}
D.~K. McClish.
\newblock {{A}nalyzing a portion of the {R}{O}{C} curve}.
\newblock \emph{Med Decis Making}, 9\penalty0 (3):\penalty0 190--195, 1989.

\bibitem[Narasimhan and Agarwal(2013{\natexlab{a}})]{narasimhan2013}
H.~Narasimhan and S.~Agarwal.
\newblock {A} structural {SVM} based approach for optimizing partial {AUC}.
\newblock In \emph{Proceedings of the 30th International Conference on Machine
  Learning}, volume JMLR W\&CP 28, pages 516--524, 2013{\natexlab{a}}.

\bibitem[Narasimhan and Agarwal(2013{\natexlab{b}})]{narasimhan2014}
H.~Narasimhan and S.~Agarwal.
\newblock {SVM$_\text{pAUC}^\text{tight}$}: A new support vector method for
  optimizing partial {AUC} based on a tight convex upper bound.
\newblock In \emph{Proceedings of the 19th ACM SIGKDD Conference on Knowledge
  Discovery and Data Mining}, pages 167--175, 2013{\natexlab{b}}.

\bibitem[Owen(2007)]{owen2007}
A.B. Owen.
\newblock Infinitely imbalanced logistic regression.
\newblock \emph{J Mach Learn Res}, 8:\penalty0 761--773, May 2007.

\bibitem[Pepe(2005)]{pepe2005}
M.~S. Pepe.
\newblock {{E}valuating technologies for classification and prediction in
  medicine}.
\newblock \emph{Stat Med}, 24\penalty0 (24):\penalty0 3687--3696, Dec 2005.

\bibitem[Pepe and Thompson(2000)]{pepe2000}
M.~S. Pepe and M.~L. Thompson.
\newblock {{C}ombining diagnostic test results to increase accuracy}.
\newblock \emph{Biostatistics}, 1\penalty0 (2):\penalty0 123--140, Jun 2000.

\bibitem[Pepe et~al.(2003)Pepe, Longton, Anderson, and Schummer]{pepe2003}
M.~S. Pepe, G.~Longton, G.~L. Anderson, and M.~Schummer.
\newblock {{S}electing differentially expressed genes from microarray
  experiments}.
\newblock \emph{Biometrics}, 59\penalty0 (1):\penalty0 133--142, Mar 2003.

\bibitem[Pepe et~al.(2006)Pepe, Cai, and Longton]{pepe2006}
M.~S. Pepe, T.~Cai, and G.~Longton.
\newblock {{C}ombining predictors for classification using the area under the
  receiver operating characteristic curve}.
\newblock \emph{Biometrics}, 62\penalty0 (1):\penalty0 221--229, Mar 2006.

\bibitem[Rudin(2009)]{rudin2009}
C.~Rudin.
\newblock {{T}he {P}-{N}orm {P}ush: a simple convex ranking algorithm that
  concentrates at the top of the list}.
\newblock \emph{JMLR}, 10:\penalty0 2233--2271, Dec 2009.

\bibitem[Schmid et~al.(2012)Schmid, Hothorn, Krause, and Rabe]{schmid2012}
M.~Schmid, T.~Hothorn, F.~Krause, and C.~Rabe.
\newblock {{A} {P}{A}{U}{C}-based estimation technique for disease
  classification and biomarker selection}.
\newblock \emph{Stat Appl Genet Mol Biol}, 11\penalty0 (5), 2012.

\bibitem[Thompson and Zucchini(1989)]{thompson1989}
M.~L. Thompson and W.~Zucchini.
\newblock {{O}n the statistical analysis of {R}{O}{C} curves}.
\newblock \emph{Stat Med}, 8\penalty0 (10):\penalty0 1277--1290, Oct 1989.

\bibitem[Tibshirani(1994)]{tibshirani1994}
R.~Tibshirani.
\newblock {{R}egression shrinkage and selection via the lasso}.
\newblock \emph{J. R. Stat. Soc. Ser. B Stat. Methodol.}, 58:\penalty0
  267--288, 1994.

\bibitem[Walter(2005)]{walter2005}
S.~D. Walter.
\newblock {{T}he partial area under the summary {R}{O}{C} curve}.
\newblock \emph{Stat Med}, 24\penalty0 (13):\penalty0 2025--2040, Jul 2005.

\bibitem[Wang and Chang(2011)]{wang2011}
Z.~Wang and Y.~C. Chang.
\newblock {{M}arker selection via maximizing the partial area under the
  {R}{O}{C} curve of linear risk scores}.
\newblock \emph{Biostatistics}, 12\penalty0 (2):\penalty0 369--385, Apr 2011.

\bibitem[Zou and Hastie(2005)]{zou2005}
H.~Zou and T.~Hastie.
\newblock {{R}egularization and variable selection via the Elastic Net}.
\newblock \emph{J. R. Stat. Soc. Ser. B Stat. Methodol.}, 67:\penalty0
  301--320, 2005.

\end{thebibliography}

\end{document}